\documentclass[preprint]{aastex}

\usepackage[]{emulateapj5}

\def\gtsim{\ {\raise-0.5ex\hbox{$\buildrel>\over\sim$}}\ }
\def\ltsim{\ {\raise-0.5ex\hbox{$\buildrel<\over\sim$}}\ }
\slugcomment{To appear in the ApJ v612, 1 September 2004}

\begin{document}

\title{The Most Exciting Massive Binary Cluster in NGC~5128:
Clues to the Formation of Globular Clusters
\thanks{Based on observations collected with the 
Very Large Telescope of the European Southern Observatory (Programme
70.B-0547(A)),
with the Magellan I Baade Telescope of the Carnegie Institution, 
and the 0.9~m Telescope of the Cerro Tololo Interamerican Observatory.
}}

\author{
 Dante Minniti $^{2}$,
 Marina Rejkuba $^{3}$,
 Jos\'e G. Funes, S.J.$^{4}$,
Robert C. Kennicutt, Jr.  $^{5}$
}

\altaffiltext{1}{Based on observations collected with the
Very Large Telescope of the European Southern Observatory (Programme
70.B-0547(A)),
with the Magellan I Baade Telescope of the Carnegie Institution,
and the 0.9~m Telescope of the Cerro Tololo Interamerican Observatory.
}

\altaffiltext{2}{Department of Astronomy, Pontificia Universidad Cat\'olica, 
Casilla 306, Santiago 22, Chile\\
E-mail:  dante@astro.puc.cl}

\altaffiltext{3}{European Southern Observatory, Karl-Schwarzschild-Str. 2, D-85748 Garching b.  M\"{u}nchen,
Germany\\
E-mail: mrejkuba@eso.org}

\altaffiltext{4}{Vatican Observatory Research Group, Steward Observatory, University of Arizona, Tucson AZ
85721, USA\\
E-mail: jfunes@as.arizona.edu}

\altaffiltext{5}{Steward Observatory, University of Arizona, Tucson AZ 85721, USA\\
E-mail: rkennicutt@as.arizona.edu}

\begin{abstract}
VLT images in $BVI$ are used to identify the ionizing source centered on Sersic 13, 
the largest HII region of the giant nearby galaxy NGC~5128 with $log L_{H\alpha}=39.6$ 
erg/s. This ionizing source turns out to be a close pair of bright and blue star 
cluster candidates.  Spectroscopy obtained with the  Magellan~I telescope confirms 
that these are massive young clusters physically associated 
with the giant HII region Sersic 13.  The spectra of both clusters show prominent 
Wolf-Rayet type emission features, and prominent lines of HI and HeI, indicative 
of a very young age ($t \approx few\times 10^6$ yr).  Their luminosities make each of 
them at least as luminous as the massive young cluster R136 in 30 Doradus in the LMC,
and their individual masses are estimated to be $1-7.5\times 10^5$ M$_{\odot}$. 
In addition, the projected separation of the cluster pair is $42$ pc.  The measured 
velocity difference between the clusters is small, $\Delta V = 49\pm 21$ km/s, and 
within 2$\sigma$ of the expected orbital velocity $V_{orb}=5-12$ km/s if they are 
bound.  Dynamical models predict that binary clusters with these properties
would merge in a short timescale of a few orbital periods ($P = 20-50\times 10^6$ yr).  
The discovery of this binary cluster suggests that mergers of young massive clusters 
could lead to the formation of the most massive globular clusters such as $\omega$Cen 
in our Galaxy and $G1$ in M31.  Alternatively, if they are not gravitationally bound,
these objects would individually evolve into two normal globular clusters.
\end{abstract}

\keywords{galaxies: individual (NGC 5128, Centaurus A) --- globular clusters: general --- galaxies: star clusters}

\section{Introduction}

The young populous clusters were defined as a special class by Hodge (1961), 
who studied blue compact clusters in the Large Magellanic Cloud (LMC).
Kennicutt \& Chu (1988) concluded that populous blue clusters may form in 
the centers of giant HII regions, such as the central massive cluster
R136 in the giant HII region 30Doradus in the LMC.  Due to its proximity,
R136 is the best studied representative of this class. The properties of the 
populous blue clusters have been discussed extensively in the literature 
(Ma\'{\i}z-Apellan\'{\i}z 2002 and references therein). They are massive, 
$M>10^5M_{\odot}$, luminous, $M_V<-10$, young, $t<10^7$ yr,
and are often embedded in giant HII regions.

For an on-going spectroscopic survey of stellar clusters in the inner regions
of the nearby giant elliptical galaxy NGC~5128, we selected candidates based on
BVI images obtained with the VLT. In particular, we targeted candidates
apparently  centered on the brightest HII regions of this galaxy, 
which have been studied by 
several authors in the past (see Phillips 1981).
The largest of these HII regions is located at the northern edge
of the dust disk of this galaxy, at
$RA(2000) = 13:25:27.5$, $DEC(2000)=-43:00:11$ with a size of about $0.7\times 1.0$ kpc.
This largest NGC~5128 HII region has been named
Nr.\ 13 by Sersic (1969), Nr.\ 13 by Moellenhoff (1981), Nr.\ 10 by 
Graham (1979),
and Nr.\ 34 by Dufour et al.\ (1979).

Even though the HII region Sersic 13 has been observed before, 
the nature of its exciting central source was not realized.  Phillips (1981) and
Rosa \& D'Odorico (1986) noticed that the WR features present were indicative
of the presence of bright young stars. Moellenhoff (1979)
reports the presence of two stellar knots in the center of this region based on 
narrow-band photography, and speculates that they are two bright O-type stars.
We extend their work here by studying the
photometric and spectroscopic properties of the components of this exciting source,
showing that it is composed of 2 very massive young populous clusters, separated by
2.25 arcsec in the N-S direction.
In order to avoid adding yet another notation, we call the clusters Sersic 13-N and
Sersic 13-S.

Typical globular clusters are made of stars formed together from
one large parent cloud. Their constituent stars share all the same 
chemical composition and age. The most
massive globular clusters, such as $\omega$Cen in the Milky Way, and
$G1$ in M31, are an exception, and their formation may have
been different from the bulk of the population.
Following the confirmation that the stellar population in these clusters
is composite (Hilker \& Richtler 2000, Pancino et al. 2000, Meylan et al. 2001), 
three different formation theories are considered for such massive 
clusters. One is the merger of binary clusters (Icke \& Alcaino 1988, 
Sugimoto \& Makino 1989, van den Bergh 1996),
the second is the stripping of nucleated dwarf 
galaxies (Freeman 1993, Meylan 2002), and the third involves self-enrichment
(e.g.\ Morgan \& Lake 1989, Parmentier et al.~1999). 

While the young binary star clusters appear to be rather frequent in smaller galaxies,
like the Magellanic Clouds (e.g.~Bhatia \& Hatzidimitriou 1988, Bhatia et al.\ 1991, 
Dieball, M\"uller \& Grebel 2002), no binary globular clusters have been found.
Either they
do not form because globular clusters generally trace kinematically hot 
populations, where the possibility of capture is rather small, or they do not
last long because close binary clusters merge rapidly after they form.
There are no known binary globular clusters in the Milky Way,
which contains 150 globular clusters in total. Perhaps binary clusters are
just rare, and one should search in a larger population.
The peculiar giant elliptical galaxy NGC~5128 (Centaurus A)
holds a large globular cluster system, $>$10 times that of the Milky Way
(Kissler-Patig 1997).
In this paper we report, for the first time,
the identification of a massive young binary star cluster pair in
NGC~5128, using images obtained with the ESO VLT, and spectroscopy
acquired with the Magellan I telescope.
The binary cluster subject of this study is a very interesting object, 
because, as we discuss below, it may lead to the formation of a very 
massive globular cluster like $\omega$Cen or G1.

The paper is organized as follows.
Section 2 gives the details of the photometric and
spectroscopic observations and reductions, and determine the 
physical properties of our targets.  In Section 3 we compare these 
targets with the young massive cluster R136 of 30Doradus in the LMC.
Section 4 discusses the possibility that the most massive
object could be a young binary globular cluster and 
Section 5 summarizes the results of this work.

\section{Photometric and Spectroscopic Properties: Ages and Velocities}

We have used 300 sec $BVI$ images taken with the European Southern Observatory 
Very Large Telescope (ESO VLT) during the commissioning of the FORS2 at UT2 to
select globular clusters in the central field of NGC~5128. Additional 
shorter exposure images amounting to 3$\times$60 sec $+$ one 20 sec exposure per
filter were obtained with FORS1 at the ESO VLT UT1 in January 11, 2003, under
photometric conditions. These were used to calibrate the FORS2 data and to
obtain the photometry of the clusters that were saturated in the longer
FORS2 exposures. The fields of view of FORS1 and FORS2 are identical,
covering the central $6\farcm8 \times 6\farcm8$ of NGC~5128.
Reductions, calibrations and globular cluster selection are discussed by
Minniti et al. (2004).
The image scale of $0\farcs2$~pix$^{-1}$ corresponds to about 0.6 pc,
projected at the distance of NGC~5128 (D=3.84 Mpc; Rejkuba 2004).
At this distance globular clusters are resolved from the ground under
subarsecond seeing conditions (Rejkuba 2001).

During that process, we noticed a number of very luminous blue sources at
the edge of the dust lane separated by a few pixels.
The brightest of these source pairs was selected for a more detailed follow-up.
H$\alpha$ images obtained with the Cerro Tololo 0.9 m telescope reveal that
this source pair is centered on Sersic 13, the largest HII region of NGC~5128.
The field is shown in Figures 1-2, where the clusters are indicated on top
of the optical and H$\alpha$ images.  The two sources have
$FWHM=3.55$ pixels in the $V$-band FORS1 images, similar to the FWHM of other
spectroscopically confirmed clusters (Harris et al.~1992) 
measured on FORS1 images, and larger than the typical stellar PSF 
of $2.9<FWHM_*<3.0$ pixels, and are also fairly round ($\epsilon = 0.2$),
suggesting that they are star clusters.

With B-V$\approx 0.4$, and $V\approx 16$, the clusters appear to be bluer and brighter
than typical old NGC~5128 globular clusters ($0.7 < B-V < 1.5$ and $18 < V < 24$) 
discussed in our previous work (Rejkuba 2001, Minniti et al. 2004).
The photometry is necessarily uncertain due to the high inhomogeneous background 
and the tightness of the pair, but it is clear that Sersic 13-N cluster
is about 0.4 magnitudes fainter in the $V$-band, and 0.2 magnitudes
bluer than the Sersic 13-S cluster.

The cluster candidates shown in Figures 1 and 2 were
observed with the Boller \& Chivens spectrograph at
the Magellan I Baade telescope on the night of 8 May 2002.
The spectra of these objects were taken 
through a 1'' wide slit, 
using a 600 line grating with 1.58~\AA/pix and 
coverage from $3800$\AA\ to $6700$\AA .  
The slit was rotated as to include both cluster candidates.
We obtained two exposures of 1200 seconds each on source
in order to eliminate cosmic ray blemishes. 
The spectral reductions and measurements were carried out in IRAF,
using the set of packages in CCDRED and TWODSPEC, as described by Minniti \& 
Rejkuba (2002).  The separation of the sources of about 2.25 arcsec combined with the
subarcsecond
seeing allowed us to extract each individual object with no contamination
from its neighbour. The positions and basic cluster data are
listed in Table 1.

Ages of clusters can be measured from photometry or spectroscopy. In fields like the
central region of NGC~5128, photometric ages are unreliable because of the uncertain
reddening corrections, and spectroscopic ages should be preferred. 

Inspection of the spectra shown in Figures 3 and 4 reveals 
that the two young star cluster candidates
have broad WR emission features (HeII 4686 and 5411, CIV 5808), as well 
as prominent HeI and HI absorption lines, and, in the case of the northern cluster,
narrow emission lines of H$\alpha$, H$\beta$, 
[OIII], [NII] and HeI.
Such spectra are indicative of very young clusters, a few million years old.

To be more precise, 
the synthetic spectral indices of Gonzalez-Delgado, Leitherer, \& Heckman (1999),
Gonzalez-Delgado \& Leitherer (1999), Leitherer (1999) allow to measure the
ages of young clusters. They list the strength of the major HI lines of the
Balmer series and HeI lines in the optical region of the spectrum
as function of age.  We use their models for an instantaneous single
burst stellar population of Solar metallicity with a Salpeter initial
mass function between 1 and 80 $M_{\odot}$.
The single burst seems to be a reasonable assumption, but we adopt
a Solar metallicity from Moellenhoff (1981).
We note that the Ca K line is stronger in the Sersic 13-N cluster, perhaps indicating
higher metallicity than the Sersic 13-S cluster (the Ca H line is blended with 
H$\epsilon$).

Table 2 lists the equivalent widths measured for HI and HeI lines for the targets. 
The absorption lines may be affected by the nebular background subtraction. 
There is an additional uncertainty in these measurements due to presence of thin cirrus 
during the night. 
By comparing with tables 1 and 2 of Gonzalez-Delgado, Leitherer, \& Heckman (1999),
we estimate for both clusters Sersic 13-N and Sersic 13-S  
ages between 1 and 10 $\times 10^6$ yr.
Emission lines in the Sersic 13-N cluster make its age estimate less accurate,
even though it is clear that this cluster is still in the nebular
phase, perhaps indicating a younger age than the Sersic 13-S cluster.
The size of the giant HII region $1.0\times 0.7$ kpc$^2$ (Figure 2)
is consistent with these ages.

We measured radial velocities using strong absorption and emission lines.
About 7 lines per cluster were measured.
The spectra bluer than 4100\AA\ were not used for the radial velocities
because the flat field and wavelength calibration are unreliable in that region.
The velocities of the Sersic 13-N and Sersic 13-S clusters are $V= 673\pm 17$ 
km/s and $624\pm 12$ km/s, respectively, which
secures their membership to NGC~5128. They also
agree with previous measurements of the
giant HII region within the errors: Moellenhoff (1981) obtains
$V = 623 \pm 35$ km/s, and Graham (1979) obtains $V = 601$ km/s.
These velocities argue in favor of the real physical 
association between the clusters and the HII regions. This was not 
granted {\it a priori} based solely on the superposition on the sky, due to the
high density of sources in the region.

\section{Comparison with the Young Cluster R136 of 30Doradus in the LMC}

The central massive cluster R136 in the giant HII region 30Doradus in the LMC is one of 
the best studied 
populous blue clusters
and it is appropriate to compare its properties
with 
those of Sersic 13-N and Sersic 13-S in NGC~5128.
The properties of R136 listed in Table 1 are from Bica et al. (1996),
Kennicutt \& Chu (1988), Bosch et al. (2001), and Ma\'{\i}z-Apellan\'{\i}z (2001).

In order to estimate the absolute luminosity of the Sersic 13-N and 13-S clusters, a reliable
measurement of the reddening must be obtained.
We chose the total reddening in the field of the HII region 
Sersic 13 from Rosa \& D'Odorico (1986)
$E(B-V)=0.23$, which is based on optical and IUE spectra.
This yields a total absorption of $A_V=0.7$ mag. For a distance of 3.84 Mpc, we obtain
$M_V=-12.4$ and $-12.8$ for the Sersic 13-N and 13-S clusters, respectively. 
For comparison, R136 has a total luminosity of $M_V=-12.4$.

Figure 4 shows the blue portion of the spectrum, containing prominent
HI and HeI absorption lines, characteristic of young objects. 
For a cluster with 1--10 Myr and $M_V=-12.4$,
Starburst 99 yields a total mass of about $1.2\times 10^5 M_{\odot}$
(doubling the mass in order to account for low mass stars). 
This mass estimate is in reasonable agreement with the mass estimated 
by comparing the luminosities with respect to R136.
For example, for the whole 30~Dor nebula a kinematic mass of $5\times 10^5$ 
M$_{\odot}$ was estimated by Bosch et al. (2001).

The total H$\alpha$ flux of $log L_{H\alpha}=39.6$ erg/s has been estimated from our 
H$\alpha$ images of this HII region (Fig. 2), without taking into account extinction.
This is comparable to the ionizing radiation of R136.

Table 3 lists all the derived physical parameters for the clusters.
We should caution that these are uncertain, and errors of 50\% in the
estimated values of masses, ages and luminosities cannot be excluded.
Table 3 shows that the clusters studied here are as extreme as or even
more extreme than R136 in 30Doradus. We can consider them as new discoveries, 
because they are not listed in the compilations
of the most massive young clusters in nearby galaxies (e.g. Ma\'{\i}z-Apellan\'{\i}z
2002). However, the physical characteristics of our targets
are comparable with the most massive of these clusters
in terms of magnitude, color, mass, age, spectral features, etc.

In addition, Portegies-Zwart et al. (2002) have analyzed Chandra  X-ray  observations
of R136, detecting a number of point sources fainter than $2\times 10^{35}$ erg/s
identified with WR and O-type stars. Minniti et al. (2004) analyzed the
optical counterparts of Chandra point sources in NGC~5128, associating some of them
with globular clusters. However, no significant emission is detected at the
position of the young massive clusters studied here. This
is not inconsistent, as the limiting flux of the Chandra observations of NGC~5128
were limited to $2\times 10^{36}$ erg/s. Deeper Chandra observations may detect
the X-ray flux in this region.

To summarize, the two clusters centered on Sersic 13, the largest HII region 
of NGC~5128, are massive clusters 
similar to R136 in 30 Doradus in the LMC. 

\section{Discussion}

The clusters Sersic 13-N and Sersic 13-S are very close in the sky, separated 
by only 2.25 arcsec in projection, and here we explore the possibility that they
form a physical binary cluster.

The Sersic 13-N cluster appears to be 0.4 mag fainter
and 0.4 mag bluer in $B-V$ than the Sersic 13-S cluster. 
Assuming that there is no differential reddening,
the magnitude and color differences could be due to
a slight difference in age, mass or chemical composition, or all.
Based on the spectra, a difference in age is preferred: the N cluster
is clearly younger because of the presence of nebular emission lines and bluer color,
and the S cluster is older because of the absence of nebular emission and 
the stronger WR features.  It is interesting to point out that if these 
clusters merge, as they age the color-magnitude diagram of the final massive
clusters may look composite (van den Bergh 1996, Catelan 1997).
However, the age difference between the clusters is small, and as the 
clusters age (i.e. when $t >10^{8}$ yr), one would no longer be able 
to detect an age difference.

In order to measure the projected separation of the clusters,
astrometry was done in the shortest exposure VLT $B$-band images, which show no
signs of saturation. By fitting Gaussian PSFs to both clusters we measure that
the cluster centers are separated by $11.2\pm 0.1$ pix. For the adopted distance 
$D=3.84\pm 0.35$ Mpc (Rejkuba 2004), this is equivalent to a projected separation 
$\Delta s= 41.7 \pm 3.8$ pc.

For a binary object, the relevant size is
the Roche radius, defined as:
$$R_R = a (0.38 + 0.2 \log m_1/m_2)^{1/2}$$ 
where {\it a} is the semimajor orbital axis, $m_1, m_2$ are the 
individual cluster masses (Paczynski 1971).
In the case of the binary cluster in NGC~5128, 
these radii would be approximately $R_R = 25.7 \pm 2.3$ pc for $a=41.7 \pm 3.8$ pc.

The Roche radius can be compared with 
the tidal radius for a cluster embedded in the potential of
a massive galaxy, as given by von Hoerners equation:
$$R_T = R (m/3M)^{1/3}$$
where R is the distance to the galaxy center, m is the cluster mass,
and M the enclosed mass of the galaxy within this distance (von Hoerner 1957). 
Based on the luminosities and ages, $m$ is taken to be $1-7.5\times 10^5 M_{\odot}$.
For NGC~5128, M can be estimated from the work of Hui et al. (1995) to be $10^{11} M_{\odot}$ 
for $R=7$ kpc (note that we use 7 kpc rather than the projected distance
to the center in the sky of 1 kpc because we assume that the clusters 
are located at the edge of the dusty disk). van Hoerners equation then yields
$R_T = 50-95$ pc,
which is much larger than the expected Roche radii of the clusters. 
Then, this can be a bound pair where the limiting radius
of each cluster will be determined by the companion cluster.

If the pair is bound, the expected orbital period in years is:
$$P_{orb} = 9.3\times 10^{7} a^{3/2} (m_1+m_2)^{-1/2}$$
where {\it a} is in pc and the masses are in $M_{\odot}$.
We estimate $P_{orb} = 20-50\times 10^6$ yr.
This orbital period is a few times the age of the clusters listed in Table 3.

The expected orbital velocity is:
$$V_{orb} = 2 \pi a/P_{orb}.$$
Using the same parameters as before, we estimate 
$V_{orb}=5-12$ km/s.
However,
this should be the maximum observed velocity difference between the clusters,
considering the projected $\sin i$ and orbital phase 
factors. The radial velocity 
difference measured here, $\Delta V = 49\pm 21$ km/s (see Table 3), is larger than
the expected orbital velocity, but a bound pair cannot be discarded within 2$\sigma$.
Note that $V_{orb} \propto (m_1+m_2)^{1/2}$, therefore a factor of 2 in both 
individual masses (our expected mass uncertainty) changes $V_{orb}$ by a factor of 2.

In summary, it is possible that the clusters Sersic 13-N and Sersic 13-S form a 
physical pair, but with the caveats of the projected $vs.$
real separation, and $sin$ $i$ factor in the measured orbital plane velocities.
However, we have to admit that the radial velocity difference between the two 
clusters is on the high side and indicate only a marginal possibility for the 
binary nature. On the other hand the uncertainties in masses and in velocities
do not exclude this possibility,
in particular in view of the recent simulations of 
Bekki et al.~(2004), who concluded that bound star clusters can be formed 
within the centers of two colliding clouds provided that the relative 
velocities of these clouds are between $10--60$ km/s. 
Measurements of more accurate velocities and
better constraint on the masses are necessary together with more sophisticated
modeling to confirm or reject the binary
nature of our target.

Several authors have discussed observations and models of binary stellar clusters 
(e.g. Icke \& Alcaino 1988, Bathia \& Hatzidimitriou 1988, Sugimoto \& Makino 1989,
Makino, Akiyama, \& Sugimoto 1991,
Gilmozzi et al. 1994, van den Bergh 1996, de Oliveira et al. 2000, 
Ballabh \& Alladin 2000, Thurl \& Johnston 2000, Dieball et al.\ 2000,
Dieball \& Grebel 2000, Dieball et al. 2002, Bekki et al.\ 2004).  
In particular, Dieball et al.~2002 made a statistical study of all 
the binary and multiple clusters
in the LMC concluding that the formation of binary clusters by tidal capture
is not likely due to low probability of close encounters of star clusters, and thus
even lower probability of tidal capture. They prefer the formation scenario proposed
by Fujimoto \& Kumai (1997), i.e.\ formation of a binary cluster from a common 
giant molecular cloud, which implies similar ages. 

Bekki et al.~(2004) investigated the formation of star clusters in a cloud-cloud 
collisions induced by tidal interaction of the LMC and the SMC. Their simulations are
important in our case, because the star formation in the central parts of NGC~5128 
could have been induced by a recent merger with a gas-rich satellite galaxy.
It is important to note that in their simulations, assuming the star 
formation efficiency of 40\% and typical masses of the colliding gas 
clouds of $10^6$~M$_\odot$, Bekki et al. (2004) create bound pairs of 
clusters with masses ($\sim 5\times 10^5$~M$_\odot$) and relative 
velocities ($10--60$ km/s) very similar to our target.
According to their simulations the impact parameter determines whether the 
two colliding clouds become a single or a binary cluster. 

Particularly relevant for our 
target are also the simulations of Sugimoto \& Makino (1989)
because of the parameters of their binary cluster models that lead to their merger. 
They perform N-body simulations of the evolution of equal mass clusters separated by a
tidal radius, finding that they merge rapidly into a massive, elliptical cluster. 
The expected merging timescale 
is just a few orbital periods, in this case $P= 20-50\times 10^6$ yr. The merging timescale is a few times larger than the 
age of our clusters ($1-10\times 10^6$ yr).
We thus suggest that the members of the exciting binary cluster Sersic 13 in 
NGC~5128 may be doomed to merge rapidly. With the additional caveat of the unknown
internal dynamical evolution of the merged system (Smith \& Gallagher 2001), 
we speculate that the final product would 
look like a more massive, flattened globular cluster, like  $\omega$Cen in 
the Milky Way, or $G1$ in M31. 

In such merger cases a bridge between clusters is expected (Dieball et al.~2000) and
has also been observed (e.g. Gurzadyan 2000,  
Dieball \& Grebel 2000). 
Figure 5 shows an expanded region around the clusters, along with the 
light contours in the $B$-band, that suggests the presence of such a bridge. 
This figure is strikingly similar to the figures of
Dieball et al. (2000) portraying the LMC cluster pair 
SL353-SL349, and the simulation of artificial clusters. 
However, the region is complex, with a number of fainter point sources and gas 
emission, and deeper imaging is desirable.

Even though the orbit of this binary star cluster is not
known, the theoretical challenge would be to simulate numerically 
the evolution in order to assess if the clusters will merge, and in what timescale.
The complementary observational challenge would be to do a complete systematic
survey of the NGC~5128 globular cluster system in order to establish
the frequency of globular cluster binarity.

\section{Conclusions}

We identify a candidate binary massive cluster in the 
inner region of NGC~5128. This pair of clusters is centered on the 
largest HII region of the galaxy, Sersic 13,
and we baptize this as the most exciting binary cluster of this galaxy.

The components have been classified as young massive clusters on the basis
of their sizes, magnitudes and colors, and the reliability of the
identification has been confirmed spectroscopically.
Both of them have Wolf-Rayet type spectra, and are at least as luminous as
R136 in the LMC.  The measured radial velocity difference 
($\Delta V = 49$ km/s), and projected separation 
($\Delta s = 42$ pc), are consistent with a binary object within the errors.

Kennicutt \& Chu (1988) suggested that giant HII regions such as 30Doradus in the
LMC can be the birth places of massive young globular clusters. 
In this paper we extend this concept, because
the discovery of this binary cluster suggests that, at 
least in some cases, mergers of young massive clusters could 
lead to the formation of the most massive globular clusters
such as $\omega$Cen in our Galaxy and $G1$ in M31.
Alternatively, if they are not gravitationally bound,
these objects would individually evolve into two normal globular clusters.
Their evolution depends on their estimated masses $>10^5 M_{\odot}$, which are
very uncertain.
Dynamical masses based on integrated high-dispersion spectroscopy are needed
to constrain the masses of these clusters in NGC~5128. 

\acknowledgments{
We are grateful to the anonymous referee for the useful suggestions.
RCK is supported by NSF grants No. AST-9900789 and AST-0307386, and
NASA grant No. NAG5-8426.
DM is supported by Fondap Center for Astrophysics No. 15010003.
We also thank Sanae Akiyama and Janice Lee for their assistance in reducing 
the $H_{\alpha}$ data presented in this paper.
 }

\clearpage
\begin{figure}
\plotone{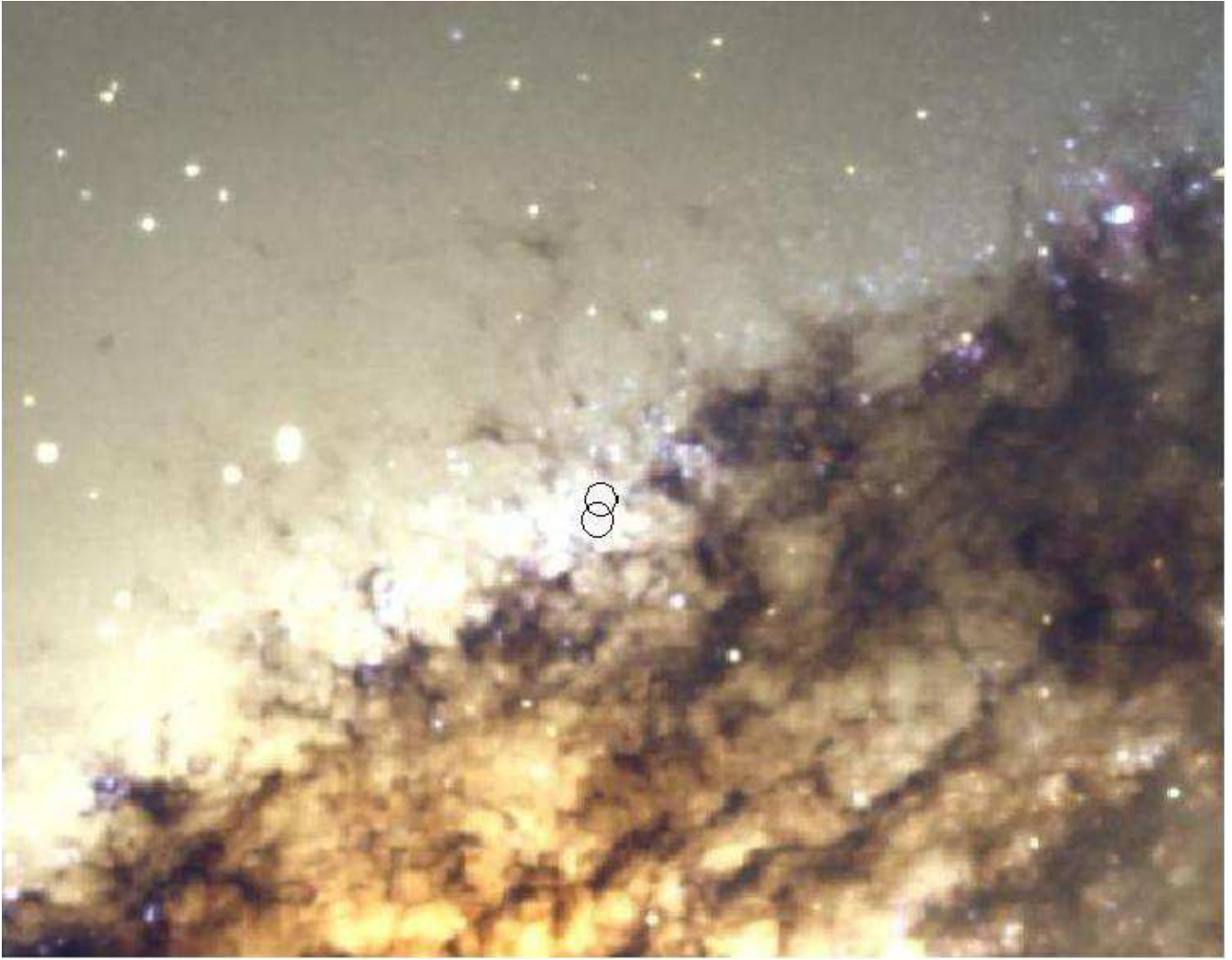}
\caption{
Location of the clusters Sersic 13-N and Sersic 13-S in NGC~5128,
in the color VLT image. The circles show the location of the clusters.
The field of view is about $2\farcm0 \times 1\farcm5$;
North is up, and East to the left.
}
\end{figure}

\begin{figure}
\plotone{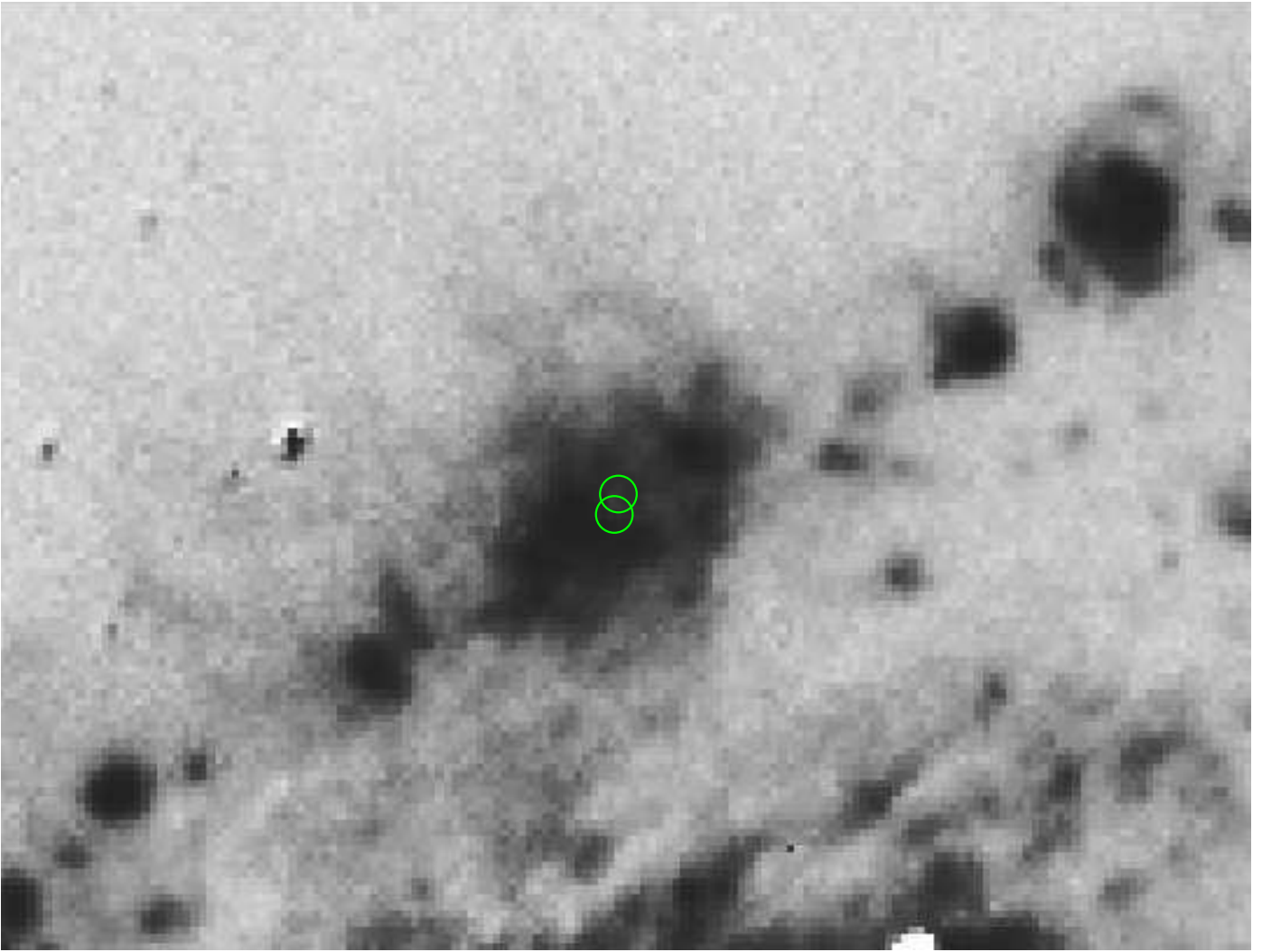}
\caption{
Same as Figure 1, showing the continuum subtracted
H$\alpha$ image of the largest HII region of NGC~5128, surrounding the
clusters Sersic 13-N and Sersic 13-S (circles).
The field of view is about $2\farcm0 \times 1\farcm5$;
North is up, and East to the left.
}
\end{figure}

\begin{figure}
\plotone{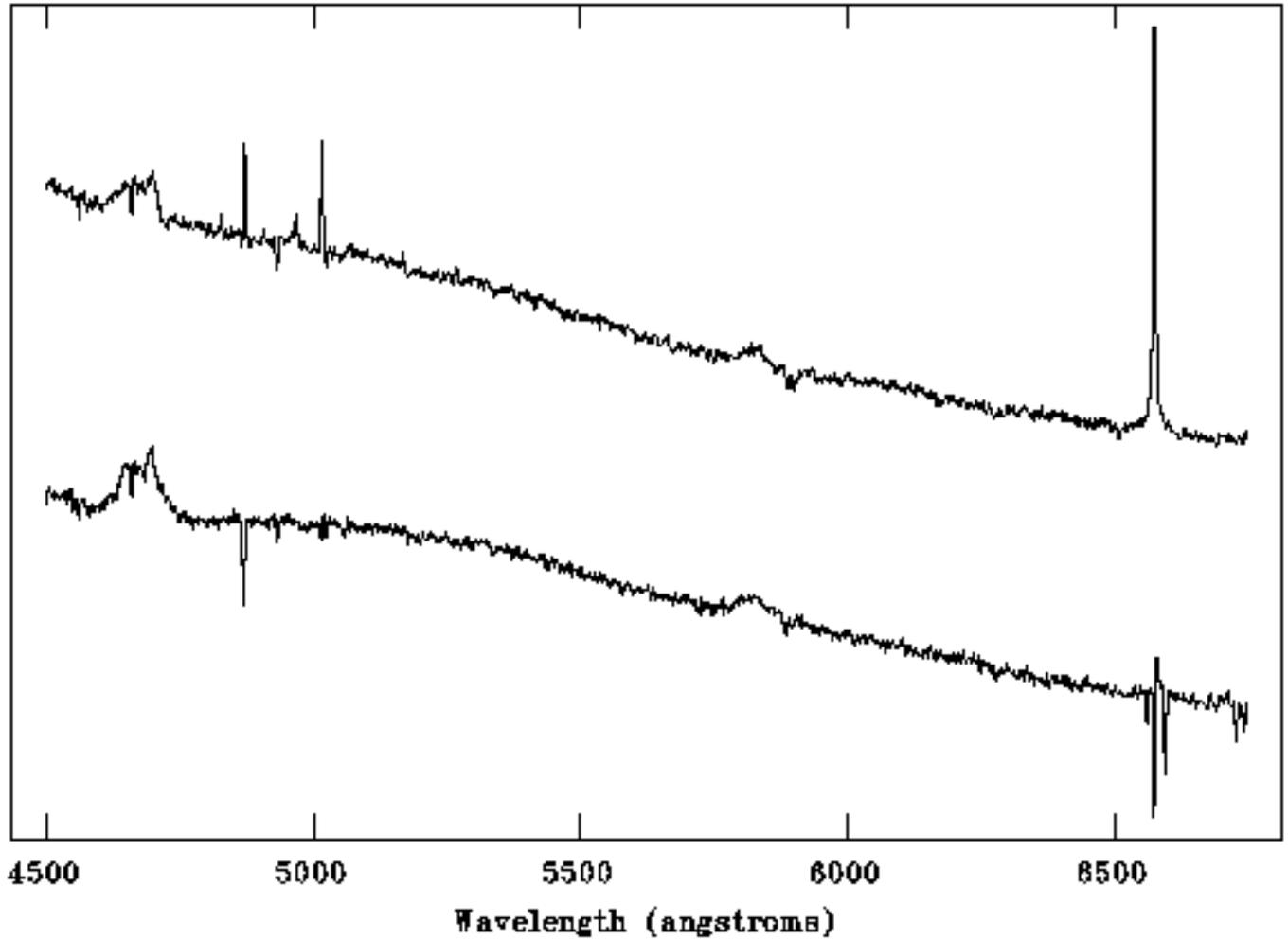}
\caption{
Red portion of the spectra of both components of 
clusters Sersic 13-N and Sersic 13-S in NGC~5128. The spectra have not been smoothed.
The northern component is at the top. The absorption lines
in the bottom spectrum are affected by local background subtraction.
}
\end{figure}

\begin{figure}
\plotone{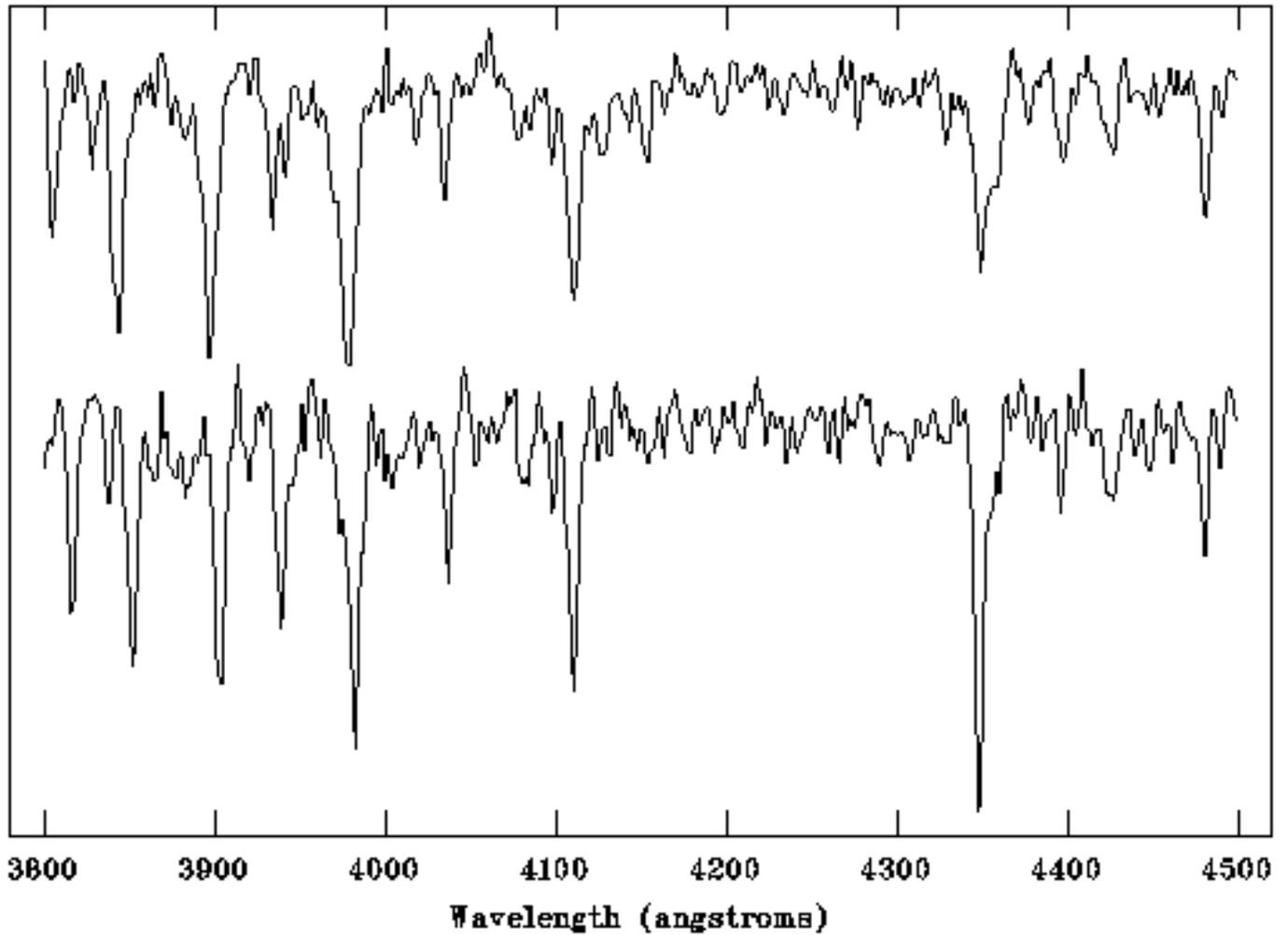}
\caption{
Blue portion of the spectra of the 
clusters Sersic 13-N and Sersic 13-S in NGC~5128.
The northern component is at the top.  The spectra have been continuum subtracted, 
but not smoothed.  We note that the shift of the
lines bluer than 4000\AA\ in the bottom spectra are an artifact: the flat-field and
wavelength calibrations are unreliable in this region.
}
\end{figure}

\begin{figure}
\plotone{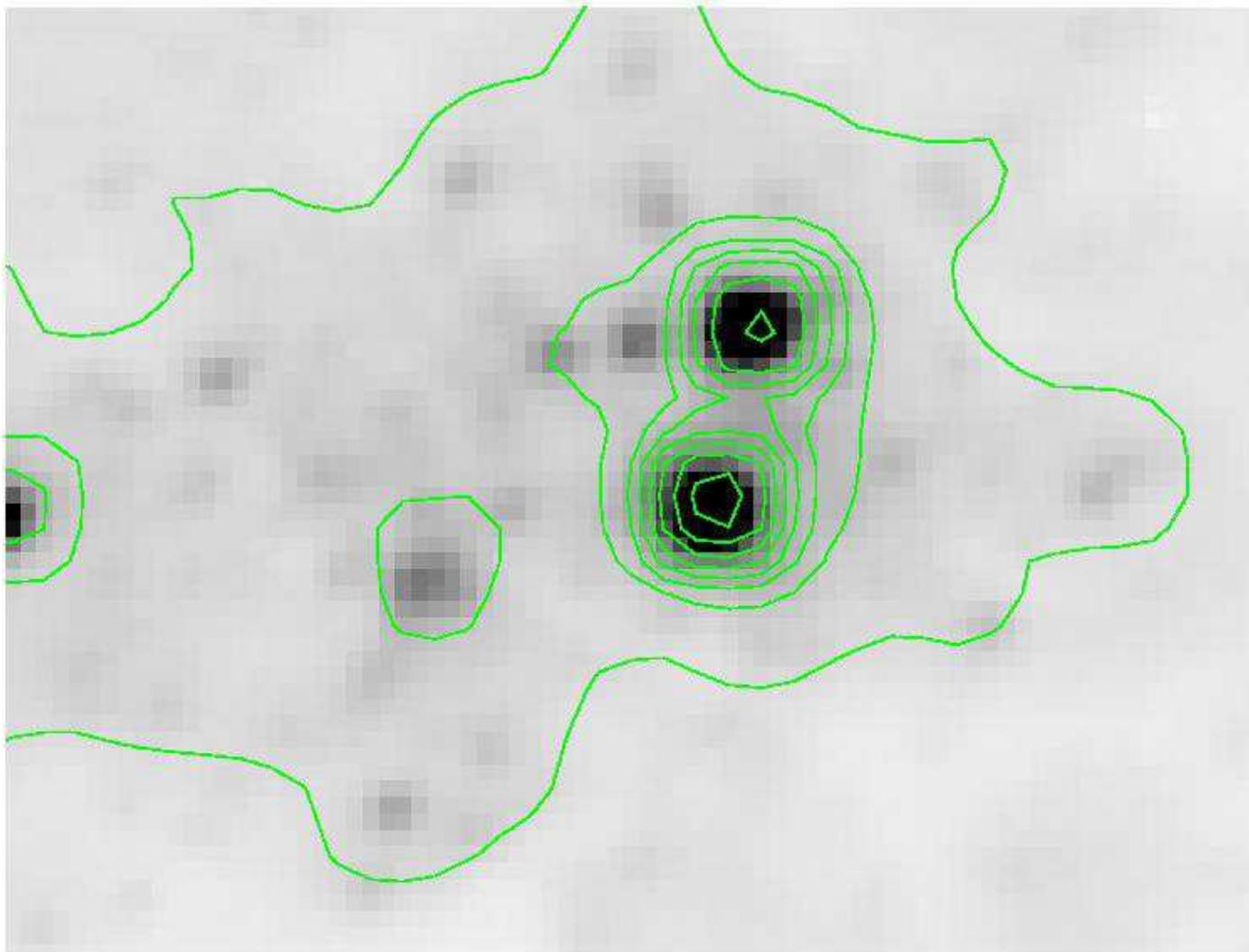}
\caption{
Expanded image of the clusters Sersic 13-N and Sersic 13-S in NGC~5128,
showing the light contours in the $B$-band. It appears as if there is a bridge
at low surface brightness between the clusters. The same is seen
in the $V$ and $I$-band images.
The projected field of view is about $230$pc $\times$ $170$ pc;
North is up, and East to the left.
}
\end{figure}

\clearpage

\begin{table}
\caption{Observed Parameters of the two Young Massive Clusters in NGC~5128}
\centerline{
\begin{tabular}{llllllll}
\hline
\hline
\multicolumn{1}{c}{ID}&
\multicolumn{1}{c}{RA}&
\multicolumn{1}{c}{DEC}&
\multicolumn{1}{c}{V}&
\multicolumn{1}{c}{B-V}&
\multicolumn{1}{c}{V-I}&
\multicolumn{1}{c}{Vr ($km s^{-1}$)}&
\multicolumn{1}{c}{FWHM}\\
\hline
Sersic 13-N&13:25:26.71&-43:00:10.4&16.2 & 0.35&0.75&624$\pm$12& 3.56 pix\\
Sersic 13-S&13:25:26.75&-43:00:12.3&15.8 & 0.55&1.15&673$\pm$17& 3.59 pix\\
R136&05:38:42.30&-69:06:03.0& 7.25&-0.02&-----&-----&-----\\
\hline
\end{tabular}
}
\end{table}

\begin{table}
\caption{Measured EWs of HI and HeI Lines}
\centerline{
\begin{tabular}{lllllllllllll}
\hline
\hline
\multicolumn{1}{c}{MBRFG}&
\multicolumn{1}{c}{H$\alpha$}&
\multicolumn{1}{c}{H$\beta$}&
\multicolumn{1}{c}{H$\gamma$}&
\multicolumn{1}{c}{H$\delta$}&
\multicolumn{1}{c}{H8}&
\multicolumn{1}{c}{H9}&
\multicolumn{1}{c}{H10}&
\multicolumn{1}{c}{He4922}&
\multicolumn{1}{c}{He4471}&
\multicolumn{1}{c}{He4388}&
\multicolumn{1}{c}{He4026}&
\multicolumn{1}{c}{He3819}\\
\hline
Sersic 13-N&-13.7&-1.6&2.6&1.8&2.3&2.1&0.6&0.8&0.9&0.9&0.7&0.6\\
Sersic 13-S&4.3&1.6&2.2&2.0&1.8&---&1.4&0.5&0.6&0.4&1.2&0.6\\
\hline
\multicolumn{13}{l}{
Negative EWs indicate emission lines.
}\\
\end{tabular}
}
\end{table}

\begin{table}
\caption{Derived Parameters of the two Young Massive Clusters in
NGC~5128}
\centerline{
\begin{tabular}{llll}
\hline
\hline
\multicolumn{1}{c}{ID}&
\multicolumn{1}{c}{M$_V$}&
\multicolumn{1}{c}{Mass}&
\multicolumn{1}{c}{Age}\\
\hline
Sersic 13-N&-12.4&$1-5\times 10^5 M_{\odot}$&1-10$\times 10^6$\\
Sersic 13-S&-12.8&$1.5-7.5\times 10^5 M_{\odot}$&1-10$\times 10^6$\\
R136&-12.4&$1-5\times 10^5 M_{\odot}$&4--5$\times 10^6$\\
\hline
\end{tabular}
}
\end{table}


\begin{references}
\reference{}{Ballabh, G. M. \& Alladin, S. M. 2000, Bull. Astr. Soc. India, 28, 261}
\reference{}{Bathia \& Hatzidimitriou 1988, MNRAS, 230, 215}
\reference{}{Bekki, K., Beasley, M.\ A., Forbes, D.\ A.\ \& Couch, W.\ J., 2004, ApJ, 602, 730}
\reference{}{Bica, E., Claria, J. J., Dottori, H., Santos, J. F. C., \& Piatti, A. E. 1996, ApJS, 102, 57}
\reference{}{Bosch, G., Selman, F., Melnick, J., \& Terlevich, R. 2001, A\&A, 380, 137}
\reference{}{Catelan, M., 1997, ApJ, 478, L99}
\reference{}{de Oliveira, M. R., Bica, E., \& Dottori, H. 2000, MNRAS, 311, 589}
\reference{}{Dieball, A.\ \& Grebel, E.\ K., 2000, A\&A, 358, 897}
\reference{}{Dieball, A., Grebel, E.\ K.\ \& Theis, C., 2000, A\&A, 358, 144}
\reference{}{Dieball, A., M\"uller, H.\ \& Grebel, E.\ K., 2002, A\&A, 391, 547}
\reference{}{Dufour, R. J., et al. 1979, AJ, 84, 284}
\reference{}{Freeman, K. C. 1993, in IAU Symp. 153, Galactic Bulges, eds. H. Dejonghe \& H.J. Habing, 
(Dordrecht: Kluwer), p. 253}
\reference{}{Fujimoto, M.\ \& Kumai, Y., 1997, AJ, 113, 249}
\reference{}{Gilmozzi, R., Kinney, E. K., Ewald, S. P., Panagia, N., \& Romaniello, M. 1994, ApJ, 435, L43}
\reference{}{Gonzalez-Delgado, R. M., \& Leitherer, C., 1999, ApJS, 125, 479}
\reference{}{Gonzalez-Delgado, R. M., Leitherer, C., \& Heckman, T. 1999, ApJS, 125, 489}
\reference{}{Graham, J. A. 1979, ApJ, 232, 60}
\reference{}{Gurzadyan, G. A. 2000, New Astronomy, 5, 349}
\reference{}{Harris, G. L. H., Geisler, D., Harris, H. C.\ \& Hesser, J. E., 1992, AJ, 104, 613}
\reference{}{Hilker, M., \& Richtler, T. 2000, A\&A, 362, 895}
\reference{}{Hodge, P.\ W., 1961, ApJ, 133, 413}
\reference{}{Hui, X., Ford, H. C., Freeman, K. C., \& Dopita, M. A 1995, ApJ, 592, 615}
\reference{}{Icke, V., \& Alcaino, G. 1988, A\&A, 204, 115}
\reference{}{Kennicutt, R. C., \& Chu, Y.-H. 1988, AJ, 95, 720}
\reference{}{Kissler-Patig, M., 1997, A\&A, 319, 83}
\reference{}{Leitherer, C. 1999, ApJS, 123, 3}
\reference{}{Ma\'{\i}z-Apellan\'{\i}z, J. 2001, ApJ, 563, 151}
\reference{}{Ma\'{\i}z-Apellan\'{\i}z, J. 
2002, in IAU Symp. 207 on "Extragalactic Star Clusters", eds. D. Geisler, E. Grebel \& D. Minniti, (ASP: San
Francisco), p. 566}
\reference{}{Makino, J., Akiyama, K., \& Sugimoto, D. 1991, Ap\&SS, 185, 63}
\reference{}{Meylan G.
2002, in IAU Symp. 207 on "Extragalactic Star Clusters", eds. D. Geisler, E. Grebel \& D. Minniti, (ASP: San
Francisco), p. 555}
\reference{}{Meylan G., et al. 2001, ApJ, 122, 830}
\reference{}{Minniti, D., \& Rejkuba, M. 2002, ApJL, 575, L59}
\reference{}{Minniti, D., Rejkuba, M., Funes, J. G. \& Akiyama, S. 2004, ApJ, in press (astro-ph/0306619)}
\reference{}{Moellenhoff, C. 1979, A\&A, 77, 141}
\reference{}{Moellenhoff, C. 1981, A\&A, 93, 248}
\reference{}{Morgan, S.\ Lake, G., 1989, ApJ, 339, 171}
\reference{}{Paczynski, B 1971, ARA\&A, 9, 183}
\reference{}{Pancino, E., Ferraro, F. R., Bellazzini, M., Piotto, M., \& Zoccali, M. 2000, ApJ, 564, L83}
\reference{}{Parmentier, G., Jehin, E., Magain, P., et al., 1999, A\&A, 352, 138}
\reference{}{Phillips, M. M. 1981, MNRAS, 197, 659}
\reference{}{Portegies Zwart, S., Pooley, D., \& Lewin, W. 2002, ApJ, 574, 762}
\reference{}{Rejkuba, M. 2001, A\&A, 369, 812}
\reference{}{Rejkuba, M. 2004, A\&A, 413, 903}
\reference{}{Rosa, M., \& D'Odorico 1986, IAU Symp. 116 on "Luminous Stars and Associations in Galaxies", eds.
C. W. H. de Loore, p. 355}
\reference{}{Sersic, J. L. 1969, Nature, 224, 253}
\reference{}{Smith, L. J.\ \& Gallagher, J. S., 2001, MNRAS, 326, 1027}
\reference{}{Sugimoto, D., \& Makino, D. 1989, PASJ, 41, 1117}
\reference{}{Thurl, C., \& Johnston, K. V. 2000, BAAS, 196, 4101}
\reference{}{van den Bergh 1996, ApJ, 471, L31}
\reference{}{von Hoerner, S. 1957, ApJ, 125, 451}
\end{references}
\end{document}